\RequirePackage{fix-cm}

\documentclass[twocolumn,epjc3]{svjour3}  
\smartqed  % flush right qed marks, e.g. at end of proof
\RequirePackage{graphicx}
\RequirePackage{amssymb}
\RequirePackage{latexsym}
\usepackage{cite}
\usepackage{booktabs}
\RequirePackage[numbers,sort&compress]{natbib}
\journalname{Eur. Phys. J. C}
\begin{document}

\title{Orbital and epicyclic frequencies around neutron and strange stars in $R^2$  gravity%\thanksref{t1}
}


\author{Kalin V. Staykov\thanksref{e1,addr1}
        \and
        Daniela D. Doneva\thanksref{e2,addr2,addr3} %etc.
        \and
        Stoytcho S. Yazadjiev\thanksref{e3,addr1,addr2}
}

\thankstext{e1}{kstaykov@phys.uni-sofia.bg}
\thankstext{e2}{daniela.doneva@uni-tuebingen.de}
\thankstext{e3}{yazad@phys.uni-sofia.bg}


\institute{Department of Theoretical Physics, Faculty of Physics, Sofia University, Sofia 1164, Bulgaria \label{addr1}
           \and
         Theoretical Astrophysics, Eberhard Karls University of T\"ubingen, T\"ubingen 72076, Germany\label{addr2}
            \and
          INRNE - Bulgarian Academy of Sciences, 1784  Sofia, Bulgaria \label{addr3}
}

\date{Received: date / Accepted: date}

\hyphenation{epi-cyclic} 

\maketitle

\begin{abstract}
According to  various models, the orbital and the epicyclic frequencies of particles moving on a circular orbit around compact objects are related to the quasi-periodic oscillations observed in the X-ray flux of some pulsars or black hole candidates. It is expected that they originate from the inner edge of the accretion discs, deep into the gravitational field of the compact objects. Considering the planned new generation X-ray timing observatories with large collective areas, the quasi-periodic oscillations might be an excellent tool for testing gravity in strong field regime and respectively alternative gravitational theories. We examine the orbital and the epicyclic frequencies of a particle moving on a circular orbit around neutron or strange stars in $R^2$ gravity. The case of slow rotation is considered too. The $R^2$ gravity results are compared to the General Relativistic case. We comment the deviations from General Relativity, as well as the deviations due to  rotation in both theories.   

\end{abstract}

\section{Introduction}
\label{intro}
In the last few years the interest in alternative theories of gravity was significantly increased. Major role for this has the experimental confirmation of the accelerated expansion of the universe and the fact that it does not fit in the predictions of General Relativity (GR)  without the introduction of the so-called dark energy, a matter with exotic properties which interacts with the visible matter only gravitationally.
A class of viable alternative theories of gravity are the so-called $f(R)$ theories. In this case we exchange the Lagrangian of the General Relativistic Einstein-Hilbet action, namely the Ricci scalar, with a more general one -- a function of the Ricci scalar, hence $f(R)$ theories ~\cite{Sotiriou2010,DeFelice2010a,Nojiri2011,Capozziello2011}.
These theories allow us to resolve the problem with the accelerating universe without the necessity of introducing such an exotic matter. However, the theory should be tested not only on cosmological but on astrophysical scales too. The predictions, concerning astrophysical effects,  should be very close to the GR ones in the weak field regime, but it is expected to deviate for strong fields.
As a source of strong gravitational fields neutron stars (NS) and black holes (BH) are natural laboratories for testing  alternative theories of gravity. In this work we are concentrating our efforts on the former.

Future X-ray timing observatories with large collecting area like SKA ~\cite{Watts2015}, NICER ~\cite{Arzoumanian2014}, LOFT ~\cite{Feroci2012}, and AXTAR ~\cite{Ray2010} may give us a chance to test gravity in strong field regime. A promising way to do that  around compact objects are the so-called quasi-periodic oscillations (QPOs). QPOs are Hz to kHz oscillations in the X-ray flux of pulsars or black hole candidates. The latter QPOs are supposed to occur only in the presence of strong gravitational fields, with their origin  in the inner edge of the accretion disc, deep into the gravitational field of the star. However, the exact source of these oscillations is unknown so there are different models for their explanation. In the beat  frequency models, some connection between the orbital frequency and the spin frequency of the central compact object is suggested. These models require some azimuthally non-uniform structure co-rotating with the central object.
In the relativistic resonance models a resonance may occur at a particular radii of the disc at which the orbital and the epicyclic frequencies have integer ratios. It is suggested that some form of resonance may occur due to the interaction between the accretion disc and the central body too. The preferred radii models suggest that  mechanism, that chooses some radii, exists. The relativistic precession models, investigated in some resent papers ~\cite{Stella1998,Stella2001,Pappas2012b, Maselli2015,Pappas2015b}, are based on the assumption that the QPOs are directly related to the orbital (Kepler) $\Omega_p$  and to the epicyclic frequencies. In the above-mentioned models all these frequencies or some of them are involved.  The radial $\nu_r$ and the vertical $\nu_{\theta}$  epicyclic frequencies occur if a particle on a stable circular orbit is perturbed. It will start to oscillate in a radial and in a vertical direction with some stable frequencies. This frequencies are the radial and the vertical epicyclic ones. A review of these models can be found in ~\cite{Klis2006}. In some papers  QPOs models  concerning oscillations of the accretion disk itself are examined too ~\cite{Rezzolla2003, Rezzolla2003a, Montero2004}.

The epicyclic frequencies of rapidly rotating strange stars are examined in the resent paper ~\cite{Gondek-Rosinska2014a}.
Some of the above mentioned models were examined in alternative theories of gravity through the years ~\cite{DeDeo2004, Vincent2014,Maselli2015, Doneva2014b}. The observed deviations from GR in these papers were quite low, except the case of rapidly rotating neutron stars in scalar-tensor theories ~\cite{Doneva2014b}.

The structure of this paper is as follow: In Section II we present the basic steps for deriving the radius of  the innermost stable circular orbit (ISCO), the orbital frequency and the radial and the vertical epicyclic frequencies. In Section III we present and discus the results for neutron and strange stars. The paper ends with  conclusions.

\section{The epicyclic frequencies of a rotating neutron star}
\label{sec:1}

In this section we briefly present the basic steps in the derivation of the equations for the radius of the innermost stable circular orbit (ISCO), the equations for the radial and for the vertical epicyclic frequencies and for the orbital frequency  ~\cite{Ryan1995,Maselli2015,Shibata1998,Pappas2012}. The equations describing stable stationary neutron star models and additional mathematical details concerning $R^2$ gravity can be found in ~\cite{Yazadjiev2014,Staykov2014,Yazadjiev2015}.

We are considering a stationary and axisymmetric spacetime with a metric

\begin{eqnarray} \label{Metric}
ds^2 = g _{tt}dt^2 + g_{rr}dr^2 + g_{\theta \theta}d\theta^2 + 2g_{t\varphi}dtd\varphi + g_{\varphi\varphi}d\varphi^2,
\end{eqnarray}
where all the metric functions depend only on the coordinates $r$ and $\theta$. The massive particles subject to the gravitational force only move on
timelike geodesics of the metric  (\ref{Metric}). The stationary and axial Killing symmetries of metric, generated by the Killing vectors $\frac{\partial}{\partial t}$ and
$\frac{\partial}{\partial \varphi}$, give rise to two constants of motion, namely $E=-u_t$ and $L=u_{\varphi}$. The first one corresponds to the energy per unit mass and the second one to the angular momentum along the axis of symmetry, and $u^{\mu} = \dot{x}^{\mu} = dx^{\mu}/d\tau$ is the four-velocity of the particle. It is not difficult one to show that the two conservation laws can be casted in the
form

\begin{eqnarray}
\frac{dt}{d\tau} = \frac{Eg_{\varphi\varphi} + Lg_{t\varphi}}{g2}, \\
\frac{d\varphi}{d\tau} = -\frac{Eg_{t\varphi} + Lg_{tt}}{g2},
\end{eqnarray}
where we defined $g2 = g_{t\varphi}^2 - g_{tt}g_{\varphi\varphi}$ for simplicity. Here $t$  denotes the coordinate time, and  $\tau$  the proper time. From  the normalization condition $g^{\mu\nu} u_{\mu}u_{\nu} = -1$, we also have

\begin{eqnarray} \label{eq:4v_norm}
g_{rr}\dot{r}^2 + g_{\theta\theta}\dot{\theta}^2 + E^2U(r,\theta) = -1
\end{eqnarray}
with

\begin{eqnarray}
U(r,\theta)= \frac{g_{\varphi\varphi} + 2l g_{t\varphi} + l^2 g_{tt}}{g2}
\end{eqnarray}
and $l=L/E$ being the proper angular momentum.The derivatives in the above equations are with respect to the proper time $\tau$.

For  $\theta=\frac{\pi}{2}$ the problem reduces to an effective one dimensional
problem
\begin{eqnarray}
\dot{r}^2 = V(r),
\end{eqnarray}
with an effective potential

\begin{eqnarray}
 V(r)=g_{rr}^{-1}\left[-1 - E^2 U(r,\theta=\frac{\pi}{2})\right]. 
\end{eqnarray}

 The stable circular orbit with a radius $\bar{r}$ is determined by the conditions $V(r_c)= 0 = V^{'}(r_c)$ and $V^{\prime\prime}(r_c)>0$, where with prime we denote the derivative with respect to $r$ . The  condition $V^{\prime\prime}(r_c)=0$ gives the ISCO radius. The angular velocity $\Omega_p$ of a particle moving on a circular equatorial orbit 
 can be found from  the geodesic equations in the following way. We write down the geodesic equations in the form
 
\begin{eqnarray}
\frac{d}{d\tau}\left(g_{\mu\nu}\frac{dx^\nu}{d\tau}\right)= \frac{1}{2}\partial_{\mu}g_{\nu\sigma} \frac{dx^\nu}{d\tau} \frac{dx^\sigma}{d\tau}
\end{eqnarray}   
which for the radial coordinate gives 

\begin{eqnarray}
\partial_{r}g_{tt} (\frac{dt}{d\tau})^2 + 2 \partial_{r}g_{t\varphi} \frac{dt}{d\tau} \frac{d\varphi}{d\tau} + \partial_{r}g_{\varphi\varphi}(\frac{d\varphi}{d\tau})^2=0 . 
\end{eqnarray} 
Taking into account that the angular velocity is defined by $\Omega_{p}=\frac{u^\varphi}{u^{t}}=\frac{d\varphi}{dt}$ we obtain from the above equation 
\begin{eqnarray}
\Omega_p = \frac{d\varphi}{dt} = \frac{-\partial_{r} g_{t\varphi} \pm \sqrt{(\partial_{r}g_{t\varphi})^2 - \partial_{r}g_{tt}\partial_{r}g_{\varphi\varphi}}}{\partial_{r}g_{rr}}.
\end{eqnarray}

To derive the epicyclic frequencies  we should  investigate small perturbations of a stable circular orbit. The perturbations are written in the form

\begin{eqnarray} \label{eq:pert_orb}
r(t) = \bar{r} + \delta r(t), \quad \theta (t) = \frac{\pi}{2} + \delta \theta(t),
\end{eqnarray}
where $\delta r(t)$ and $\delta \theta(t)$ are perturbations to the stable circular orbit with coordinate radius $\bar{r}$ in the equatorial plane. The perturbations could be written explicitly in the form $\delta r(t) \sim e^{2\pi i \nu_{r}t} $ and $\delta \theta(t) \sim   e^{2\pi i \nu_{\theta}t}$. Substituting (\ref{eq:pert_orb}) in eq. (\ref{eq:4v_norm}) and after some calculations, and a change of the proper time $\tau$ with the coordinate one $t$, we obtain the expressions for the radial and the vertical epicyclic frequencies:

\begin{eqnarray}
\nu_{r}^{2} = \frac{\left(g_{tt} + \Omega_p g_{t\varphi}\right)^2}{2\left(2\pi\right)^2 g_{rr}} \partial_{r}^{2} U\left(\bar{r},\frac{\pi}{2}\right), \\
\nu_{\theta}^{2} = \frac{\left(g_{tt} + \Omega_p g_{t\varphi}\right)^2}{2\left(2\pi\right)^2 g_{\theta\theta}} \partial_{\theta}^{2} U\left(\bar{r},\frac{\pi}{2}\right).
\end{eqnarray}

For the case of static neutron stars the orbital frequency and the vertical epicyclic frequency coincide, i.e. $\nu_{\theta} = \nu_p$, for $f = 0$, where $f$ is the rotational frequency of the star ($f = \frac{\Omega}{2 \pi}$).
At the ISCO the square of the radial epicyclic frequency is equal to zero, and for smaller radius it is negative, which shows a radial instabilities for orbits with radius smaller than the ISCO.

For simple accretion disc models the inner edge of the disc is defined by the ISCO. For models with smaller masses the ISCO is in the interior of the star and for massive ones the radius of the ISCO is bigger than the radius of the star itself. The inner edge of the accretion disc, therefore, can reach down to the surface of the star in the former case and  to the ISCO in the latter one.

\section{Numerical results}
\label{sec:2}

We investigate what are the changes in the radius of the ISCO and in the orbital and epicyclic frequencies in $f(R)$ gravity with Lagrangian $f(R) = R + aR^2$, the so-called $R^2 $ gravity. The results are compared to  pure GR. Deviations due to rotation, in slow rotation approximation, are examined too. We consider two hadronic equations of state (EOS) and a quark one. For the hadronic EOS we are using peacewise polytropic approximation ~\cite{Read2009}. The ones we choose are APR4 and MS1.  APR4 has maximal mass not much bigger than the observational limit of two solar masses (medium stiffness), and MS1 is a stiffer one with  higher masses and radii. We examined a soft EOS too but the results are qualitatively the same, and because of that, graphs are not presented. The quark EOS has the analytical form

\begin{equation}
p = b(\rho - \rho_0),
\end{equation}
where the constants $b$ and $\rho_0$ are taken from \cite{Gondek-Rosinska2008} for EOS SQS B60. This EOS leads to maximal masses slightly bellow two solar masses, but we find it to be a good representative.

In this section we examine static, $f = 0$ Hz, and slowly rotating, $f = 80$ Hz and $ f = 160$ Hz, models of neutron and strange stars. First we will discus the deviations from pure GR due to $R^2$ gravity and as a next step we will examine the differences of the results for different rotational frequencies. The presented results in all figures are up to the maximal mass. For the presented results in  $R^2$ gravity, the maximal deviation from GR is for the maximal adopted value of the parameter $a = 10^4$.  This number is close to the maximal values allowed by the observations  $a \sim 10^5$ or in dimensional units -- $a \lesssim 5 \times 10^{11} m^2$ ~\cite{Naef2010}. 

\begin{figure*}[]
	\centering
	\includegraphics[width=1\textwidth]{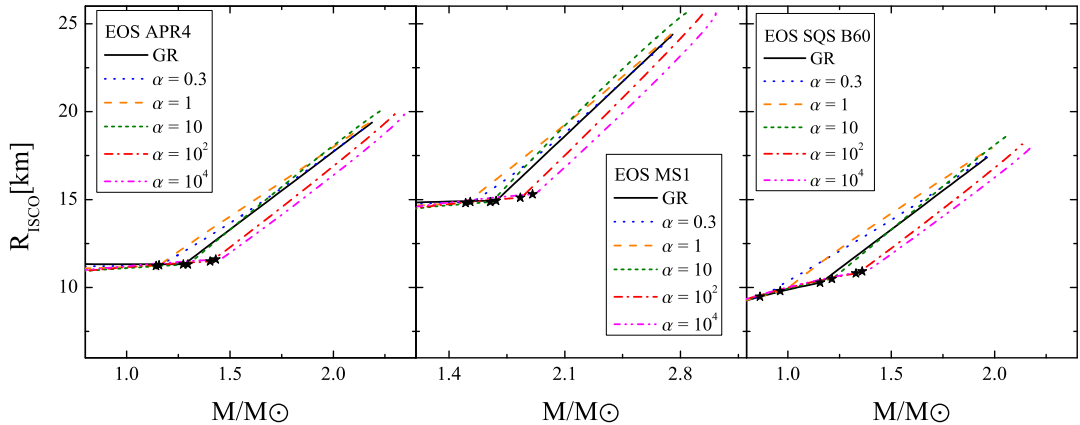}
	\caption{The radius of the ISCO in km as a function of the stellar mass  for different values of the parameter $a$. In the left panel  the results for EOS APR4 are plotted, in the middle one for EOS MS1, and in the right one for SQS B60. The results are for static neutron stars (f = 0 Hz). }
	\label{Fig:ISCO}
\end{figure*}

In Fig. \ref{Fig:ISCO} sequences of models, representing the radius of the ISCO as a function of the stellar mass are plotted. A wide range of values for the parameter $a$ is examined. In the left panel the results for EOS APR4 are plotted, in the middle one for EOS MS1, and in the right one for EOS SQS B60.  If the ISCO is in the interior of the star, that happens for models with low masses, we are plotting the radius of the star instead. The points in the different sequences of models, where the radius of the ISCO gets equal to  the radius of the star are marked with an asterisk. For all examined EOS the results are qualitatively the same. For models with  maximal masses the largest deviation is for the maximal values of the parameter $a$.  It is around 10 \% and decreases with the decrease of $a$. In the limiting case of $a \rightarrow 0 $ the solutions converge to the GR ones.

At this point a comment concerning the presented results  should be done. In Fig. \ref{Fig:ISCO}, in the present paper,  the following behaviour for the models for which the radius of the  ISCO gets bigger than the radius of the star  can be observed. Let us start the discussion with  the models marked with an asterisk. For small values of the parameter $a$ they have smaller radius and lower mass than the GR case. The $a = 10$ models have parameters which are  close to the GR ones and for high values of $a$ the radii and the masses  are higher compared to the GR case. When the ISCO is outside the star its radius  is bigger compared to the GR case for models with equal masses and small values of $a$. For $a = 10$ the results are comparable to the GR ones. For bigger values of $a$ the radius of the ISCO is smaller than the GR one. The explanation of this behaviour we find in the non-monotonous behaviour of some stellar parameters as a function of $a$. This is demonstrated, for example, in Fig. 3 in \cite{Yazadjiev2014}  for the maximal mass. It was shown that for small values of $a$ the mass decreases compared the GR one. At some value of the parameter $a$ there is a turning point and the mass starts to increase, reaching maximal masses higher than the GR one. Our investigations show that this is the case for neutron stars as well as for strange stars. Such behavior can be observed also for other stellar parameters, although in the different cases the minimum will be shifted to some other values of the parameter $a$.   If one, for example, calculates the mass and the radius for models with fixed central energy density in GR and in $R^2$ gravity, the same pattern can be found.  It is not hard to see that there is a similar dependence for the models marked with an asterisk as well as for the case when the ISCO is outside the star. This leads us to the conclusion that this behaviour is characteristic one for the theory. Naturally, similar patterns can be found in the other graphs presented below.   

\begin{figure*}[]
	\centering
	\includegraphics[width=1\textwidth]{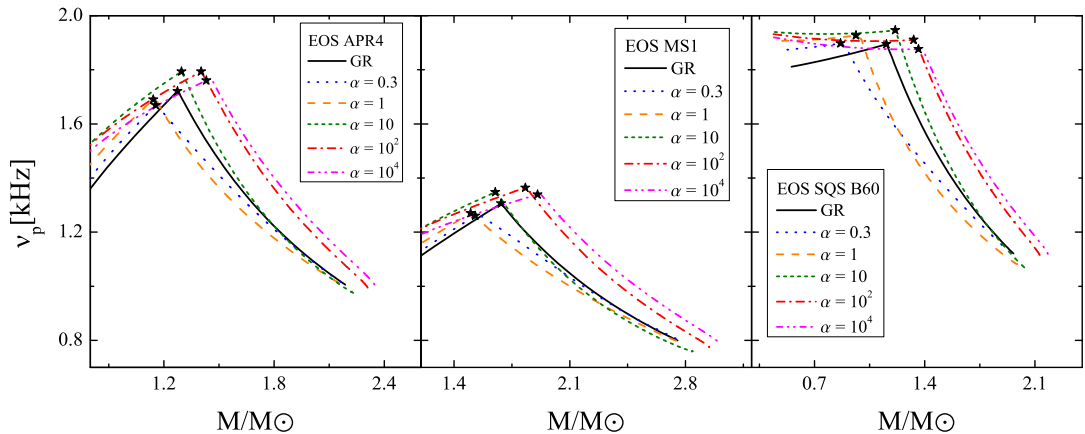}
	\caption{The orbital frequency of a particle on a circular orbit in kHz as a function of the mass of the star for different values of the parameter $a$. In the left panel  the results for EOS APR4 are plotted, in the middle one for EOS MS1, and in the right one for SQS B60. The results are for static neutron stars (f = 0 Hz). }
	\label{Fig:nu_p}
\end{figure*}

In Fig. \ref{Fig:nu_p} we plot the orbital frequency $\nu_p = \Omega_p/2\pi$, in kHz, at the ISCO as a function of the stellar mass.  If the  ISCO is in the interior of the star,  we calculated $\nu_p$ at the surface of the star. For all EOS we see qualitatively the same results. The deviations from GR for models with $a = 10^4$ is in average around 15 \% and it does not change significantly with the  mass. For models with small value of $a$ the deviation from GR, when the radius of the ISCO get bigger than  the radius of the star, is close to the maximal one, but it rapidly decreases with the increase of the mass, and converge to GR for maximal masses.

\begin{figure*}[]
	\centering
	\includegraphics[width=1\textwidth]{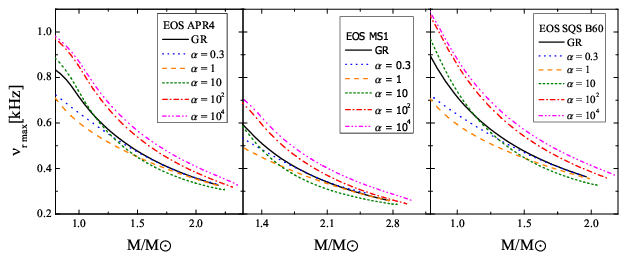}
	\caption{The maximal value of the radial epicyclic frequency in kHz as a function of the stellar mass  for different values of the parameter $a$. In the left panel  the results for EOS APR4 are plotted, in the middle one for EOS MS1, and in the right one for SQS B60. The results are for static neutron stars (f = 0 Hz). }
	\label{Fig:nu_r_max}
\end{figure*}

In Fig. \ref{Fig:nu_r_max} we plot the maximal value of the radial epicyclic frequency, in kHz, as a function of the stellar mass. 
If the ISCO is in the interior of the star, we take the maximal value on the surface or outside the star.   In the examined interval of masses the frequency monotonically decreases. For higher values of the parameter $a$ the maximal radial epicyclic frequency is higher, compared to the GR one.
For small values of $a$ the frequencies are lower than the GR ones. For smaller masses there are higher deviations, but for maximal masses the plots converge to GR.

The deviations due to rotation are qualitatively and quantitatively the same for the hadronic and for the quark EOS. This, combined with the small magnitude of the deviations is the reason why we chose  to include only the favoured by the observations hadronc EOS in the following graphs.

\begin{figure*}[]
	\centering
	\includegraphics[width=0.6\textwidth]{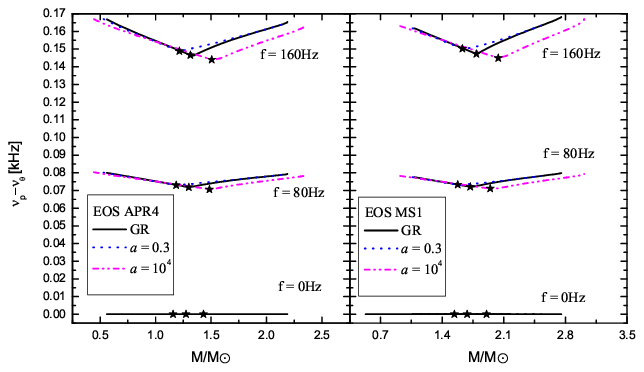}
	\caption{The difference between the orbital and the vertical epicyclic frequency at the ISCO. The plots are for different rotational rates for GR and for $R^2$ gravity with $a = 0.3$ and $a = 10^4$. ln the left panel the results for EOS APR4 are plotted, and in  the right one for EOS MS1. }
	\label{Fig:nu_p_theta}
\end{figure*}

In Fig. \ref{Fig:nu_p_theta} we plot the difference between the orbital frequency $\nu_p$ and the vertical epicyclic frequency $\nu_{\theta}$, i.e. the nodal precession frequency $\nu_n$,  as a function of the mass. In the left panel we plot the results for EOS APR4 and in the right panel for  EOS MS1. The plots are for three different rotational rates: $f = 0 $ Hz (the static case), $f = 80$ Hz, and $f = 160$ Hz. In black continuous lines are the result for GR and in colour dashed line -- the results for $a = 0.3$ and $a = 10^4$.
As we said before,  $\nu_p $ and $ \nu_{\theta}$ coincide in the static case. With the increase of the rotational frequency of the star the differences between the two frequencies increase.The difference is calculated on the surface of the star if the ISCO is inside the star, and at the ISCO in the opposite case. The transition models are marked with asterisks, in consistence with the previous graphs. The nodal precession frequency tends to decrease with the increase of the mass for models having  ISCO in the interior of the star. It increases with the increase of the mass for models having  ISCO outside the star.
In the case of $R^2$ gravity we plot the results for $a = 10^4$ and $a = 0.3$. The former value of the parameter gives the maximal deviation from GR, and the latter we choose because of the interesting behaviour demonstrated in the graphs above. If the ISCO is in the interior of the star, the results are the same for all rotational rates. In the opposite case, when the ISCO is  outside the star, the behaviour is the following. For $a = 0.3$, $\nu_n$ is slightly higher than the GR case, and rapidly converge to GR with the increase of the mass. For $a = 10^4$, $\nu_n$ has lower values than the GR one. The deviation from GR is more or less constant with the increase of the mass, and it is about 4 \%.  

\begin{figure*}[]
	\centering
	\includegraphics[width=0.6\textwidth]{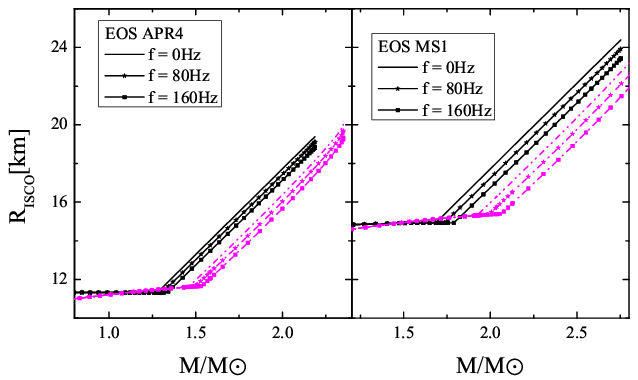}
	\caption{The radius of the ISCO, in km, as a function of the stellar mass. The plots are for GR  and for $R^2$ gravity with $a = 10^4$  and for stars rotating with $f = 0$ Hz, $f = 80$ Hz, and $f = 160 $ Hz. ln the left panel the results for EOS APR4 are plotted, and in  the right one for EOS MS1.  }
	\label{Fig:ISCO_f}
\end{figure*}

\begin{figure*}[]
	\centering
	\includegraphics[width=0.6\textwidth]{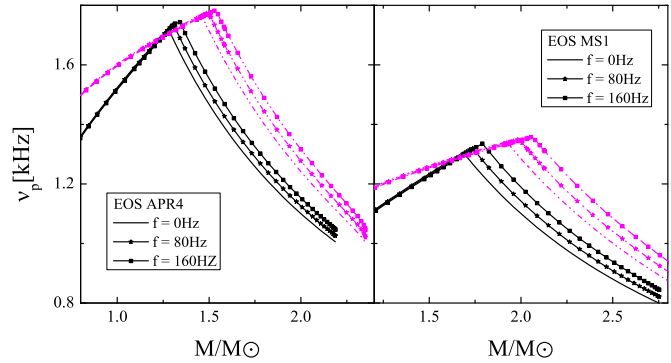}
	\caption{The orbital frequency at the  ISCO, in kHz, as function of the stellar mass. The plots are for GR  and for $R^2$ gravity with $a = 10^4$  for stars rotating with $f = 0$ Hz, $f = 80$ Hz, and $f = 160 $ Hz. ln the left panel the results for EOS APR4 are plotted, and in  the right one for EOS MS1.  }
	\label{Fig:nu_p_f}
\end{figure*}

\begin{figure*}[]
	\centering
	\includegraphics[width=0.6\textwidth]{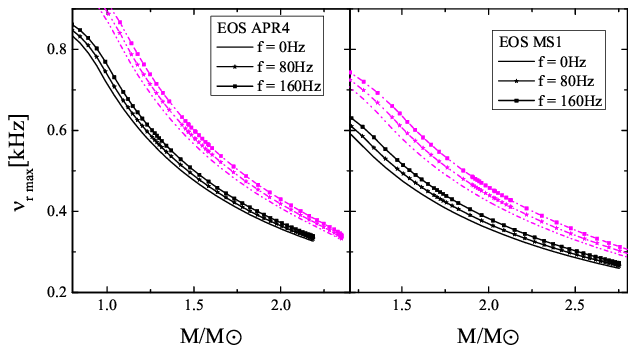}
	\caption{The maximal radial epicyclic frequency as a function of the stellar mass. The plots are for GR  and for $R^2$ gravity with $a = 10^4$  for stars rotating with $f = 0$ Hz, $f = 80$ Hz, and $f = 160 $ Hz. ln the left panel the results for EOS APR4 are plotted, and in  the right one for EOS MS1. }
	\label{Fig:nu_r_max_f}
\end{figure*}

In Figs. \ref{Fig:ISCO_f}, \ref{Fig:nu_p_f}, \ref{Fig:nu_r_max_f} we plot the radius of the ISCO, the orbital frequency and the maximal radial epicyclic frequency as a function of the mass in the case of GR and the maximal deviation in $R^2$ gravity ($a = 10^4$)  for the aforementioned  values of $f$. The rotation have the same effect for all these quantities in GR and in $R^2$ gravity. The reason why we are skipping the interestingly behaving case of $a=0.3$ is the following. The changes of the ISCO and the examined frequencies for different rotational rates causes the graphs for $a = 0.3$ and $a = 1$ to overlap with the GR ones. We find  that case to be quite uninformative so only the maximal deviation is presented, namely $a = 10^4$. Due to  rotation the radius of the ISCO decreases with about 4 \%, and the frequencies increase  with the same magnitude. The percentage deviations are the same for  GR and for $R^2$ gravity.

\section{Conclusions}
\label{sec:3}

In this paper we examined the orbital and the epicyclic frequencies of  particles moving in circular orbits around neutron and strange stars.  
Various models relate all these frequencies or some of them to QPOs. Taking into consideration the new generation observatories for timing of X-ray pulsars  with large collective area which are under construction now, QPOs are expected to have major role in testing strong field regime of gravity.

$ R^2$ gravity is a viable alternative to GR. We examined how changing the parameter of the theory effects the radius of the ISCO, the orbital and the epicyclic frequencies. Most of the observed neutron stars are slowly rotating, so we concentrated our efforts on the static and slowly rotating solutions. We investigated the deviations of the slow rotating solutions from the static case too.

We concerned two hadronic and one quark equations of state and for both cases we observed qualitatively the same behaviour with the change of the parameter of the theory. The largest deviations from GR occurs for the maximal adopted value of the parameter $a$. The radius of the ISCO decreases with about 10\% and the orbital and the maximal radial epicyclic frequencies increases with 15 - 20 \%. However, the effect due to  slow rotation is expectedly small. For the highest examined rotational frequency ($f = 160$ Hz) the deviation from the static case is about 4\% for all examined quantities.

The deviations  for slowly rotating stars, up to $160$ Hz, are much smaller than the deviations due to the $R^2$ gravity. Because of that static theoretical and numerical models can be used for comparison with observational data even for slowly rotating stars. 

The frequencies examined in this paper provide us with an opportunity to test GR in strong field regime, as well as alternative theories. Results from the numerical models could be compared to the observational data expected from the new generation X-Ray timing observatories. However, even if the observational data do not provide  a concrete information about the  correct gravitational theory, it will provide us with an unique opportunity to restrict the wide range of possible values for the free parameter of the theory.

\begin{acknowledgements}
DD would like to thank the European Social Fund and the Ministry Of Science, Research and the Arts Baden-W\"urttemberg for the support. KS and  SY would like to thank the Research Group Linkage Programme of the Alexander von Humboldt Foundation for the support. The support by the Bulgarian NSF Grant DFNI T02/6, Sofia University Research Fund under Grant 70/2015 and "New-CompStar" COST Action MP1304 is gratefully acknowledged.
\end{acknowledgements}

\bibliographystyle{spphys}       % APS-like style for physics
\bibliography{QPO}

\begin{thebibliography}{10}
\providecommand{\url}[1]{{#1}}
\providecommand{\urlprefix}{URL }
\expandafter\ifx\csname urlstyle\endcsname\relax
  \providecommand{\doi}[1]{DOI \discretionary{}{}{}#1}\else
  \providecommand{\doi}{DOI \discretionary{}{}{}\begingroup
  \urlstyle{rm}\Url}\fi

\bibitem{Sotiriou2010}
T.P. Sotiriou, V.~Faraoni, Rev.Mod.Phys. \textbf{82}, 451 (2010).
\newblock \doi{10.1103/RevModPhys.82.451}

\bibitem{DeFelice2010a}
A.~De~Felice, S.~Tsujikawa, Living Rev.Rel. \textbf{13}, 3 (2010).
\newblock \doi{10.12942/lrr-2010-3}

\bibitem{Nojiri2011}
S.~Nojiri, S.D. Odintsov, Phys.Rept. \textbf{505}, 59 (2011).
\newblock \doi{10.1016/j.physrep.2011.04.001}

\bibitem{Capozziello2011}
S.~{Capozziello}, M.~{de Laurentis}, Phys. Rep. \textbf{509}, 167 (2011).
\newblock \doi{10.1016/j.physrep.2011.09.003}

\bibitem{Watts2015}
A.~Watts, R.~Xu, C.~Espinoza, N.~Andersson, J.~Antoniadis, D.~Antonopoulou,
  S.~Buchner, S.~Dai, P.~Demorest, P.~Freire, J.~Hessels, J.~Margueron,
  M.~Oertel, A.~Patruno, A.~Possenti, S.~Ransom, I.~Stairs, B.~Stappers,
  (2015)

\bibitem{Arzoumanian2014}
Z.~{Arzoumanian}, K.C. {Gendreau}, C.L. {Baker}, T.~{Cazeau}, P.~{Hestnes},
  J.W. {Kellogg}, S.J. {Kenyon}, R.P. {Kozon}, K.C. {Liu}, S.S.
  {Manthripragada}, C.B. {Markwardt}, A.L. {Mitchell}, J.W. {Mitchell}, C.A.
  {Monroe}, T.~{Okajima}, S.E. {Pollard}, D.F. {Powers}, B.J. {Savadkin}, L.B.
  {Winternitz}, P.T. {Chen}, M.R. {Wright}, R.~{Foster}, G.~{Prigozhin},
  R.~{Remillard}, J.~{Doty}, in \emph{Society of Photo-Optical Instrumentation
  Engineers (SPIE) Conference Series}, \emph{Society of Photo-Optical
  Instrumentation Engineers (SPIE) Conference Series}, vol. 9144 (2014),
  \emph{Society of Photo-Optical Instrumentation Engineers (SPIE) Conference
  Series}, vol. 9144, p.~20.
\newblock \doi{10.1117/12.2056811}

\bibitem{Feroci2012}
M.~{Feroci}, L.~{Stella}, M.~{van der Klis}, T.J.L. {Courvoisier},
  M.~{Hernanz}, R.~{Hudec}, A.~{Santangelo}, D.~{Walton}, A.~{Zdziarski},
  D.~{Barret}, T.~{Belloni}, J.~{Braga}, S.~{Brandt}, C.~{Budtz-J{\o}rgensen},
  S.~{Campana}, J.W. {den Herder}, J.~{Huovelin}, G.L. {Israel}, M.~{Pohl},
  P.~{Ray}, A.~{Vacchi}, S.~{Zane}, A.~{Argan}, P.~{Attin{\`a}},
  G.~{Bertuccio}, E.~{Bozzo}, R.~{Campana}, D.~{Chakrabarty}, E.~{Costa},
  A.~{De Rosa}, E.~{Del Monte}, S.~{Di Cosimo}, I.~{Donnarumma},
  Y.~{Evangelista}, D.~{Haas}, P.~{Jonker}, S.~{Korpela}, C.~{Labanti},
  P.~{Malcovati}, R.~{Mignani}, F.~{Muleri}, M.~{Rapisarda}, A.~{Rashevsky},
  N.~{Rea}, A.~{Rubini}, C.~{Tenzer}, C.~{Wilson-Hodge}, B.~{Winter},
  K.~{Wood}, G.~{Zampa}, N.~{Zampa}, M.A. {Abramowicz}, M.A. {Alpar},
  D.~{Altamirano}, J.M. {Alvarez}, L.~{Amati}, C.~{Amoros}, L.A. {Antonelli},
  R.~{Artigue}, P.~{Azzarello}, M.~{Bachetti}, G.~{Baldazzi}, M.~{Barbera},
  C.~{Barbieri}, S.~{Basa}, A.~{Baykal}, R.~{Belmont}, L.~{Boirin},
  V.~{Bonvicini}, L.~{Burderi}, M.~{Bursa}, C.~{Cabanac}, E.~{Cackett}, G.A.
  {Caliandro}, P.~{Casella}, S.~{Chaty}, J.~{Chenevez}, M.J. {Coe},
  A.~{Collura}, A.~{Corongiu}, S.~{Covino}, G.~{Cusumano}, F.~{D'Amico},
  S.~{Dall'Osso}, D.~{De Martino}, G.~{De Paris}, G.~{Di Persio}, T.~{Di
  Salvo}, C.~{Done}, M.~{Dov{\v c}iak}, A.~{Drago}, U.~{Ertan}, S.~{Fabiani},
  M.~{Falanga}, R.~{Fender}, P.~{Ferrando}, D.~{Della Monica Ferreira},
  G.~{Fraser}, F.~{Frontera}, F.~{Fuschino}, J.L. {Galvez}, P.~{Gandhi},
  P.~{Giommi}, O.~{Godet}, E.~{G{\"o}{\v g}{\"u}{\c s}}, A.~{Goldwurm},
  D.~{G{\"o}tz}, M.~{Grassi}, P.~{Guttridge}, P.~{Hakala}, G.~{Henri},
  W.~{Hermsen}, J.~{Horak}, A.~{Hornstrup}, J.J.M. {in't Zand}, J.~{Isern},
  E.~{Kalemci}, G.~{Kanbach}, V.~{Karas}, D.~{Kataria}, T.~{Kennedy},
  D.~{Klochkov}, W.~{Klu{\'z}niak}, K.~{Kokkotas}, I.~{Kreykenbohm},
  J.~{Krolik}, L.~{Kuiper}, I.~{Kuvvetli}, N.~{Kylafis}, J.M. {Lattimer},
  F.~{Lazzarotto}, D.~{Leahy}, F.~{Lebrun}, D.~{Lin}, N.~{Lund},
  T.~{Maccarone}, J.~{Malzac}, M.~{Marisaldi}, A.~{Martindale},
  M.~{Mastropietro}, J.~{McClintock}, I.~{McHardy}, M.~{Mendez},
  S.~{Mereghetti}, M.C. {Miller}, T.~{Mineo}, E.~{Morelli}, S.~{Morsink},
  C.~{Motch}, S.~{Motta}, T.~{Mu{\~n}oz-Darias}, G.~{Naletto}, V.~{Neustroev},
  J.~{Nevalainen}, J.F. {Olive}, M.~{Orio}, M.~{Orlandini}, P.~{Orleanski},
  F.~{Ozel}, L.~{Pacciani}, S.~{Paltani}, I.~{Papadakis}, A.~{Papitto},
  A.~{Patruno}, A.~{Pellizzoni}, V.~{Petr{\'a}{\v c}ek}, J.~{Petri}, P.O.
  {Petrucci}, B.~{Phlips}, L.~{Picolli}, A.~{Possenti}, D.~{Psaltis},
  D.~{Rambaud}, P.~{Reig}, R.~{Remillard}, J.~{Rodriguez}, P.~{Romano},
  M.~{Romanova}, T.~{Schanz}, C.~{Schmid}, A.~{Segreto}, A.~{Shearer},
  A.~{Smith}, P.J. {Smith}, P.~{Soffitta}, N.~{Stergioulas}, M.~{Stolarski},
  Z.~{Stuchlik}, A.~{Tiengo}, D.~{Torres}, G.~{T{\"o}r{\"o}k}, R.~{Turolla},
  P.~{Uttley}, S.~{Vaughan}, S.~{Vercellone}, R.~{Waters}, A.~{Watts},
  R.~{Wawrzaszek}, N.~{Webb}, J.~{Wilms}, L.~{Zampieri}, A.~{Zezas},
  J.~{Ziolkowski}, Experimental Astronomy \textbf{34}, 415 (2012).
\newblock \doi{10.1007/s10686-011-9237-2}

\bibitem{Ray2010}
P.S. {Ray}, D.~{Chakrabarty}, C.A. {Wilson-Hodge}, B.F. {Phlips}, R.A.
  {Remillard}, A.M. {Levine}, K.S. {Wood}, M.T. {Wolff}, C.S. {Gwon}, T.E.
  {Strohmayer}, M.~{Baysinger}, M.S. {Briggs}, P.~{Capizzo}, L.~{Fabisinski},
  R.C. {Hopkins}, L.S. {Hornsby}, L.~{Johnson}, C.D. {Maples}, J.H. {Miernik},
  D.~{Thomas}, G.~{de Geronimo}, in \emph{Society of Photo-Optical
  Instrumentation Engineers (SPIE) Conference Series}, \emph{Society of
  Photo-Optical Instrumentation Engineers (SPIE) Conference Series}, vol. 7732
  (2010), \emph{Society of Photo-Optical Instrumentation Engineers (SPIE)
  Conference Series}, vol. 7732, p.~48.
\newblock \doi{10.1117/12.857385}

\bibitem{Stella1998}
L.~{Stella}, M.~{Vietri}, ApJ \textbf{492}, L59 (1998).
\newblock \doi{10.1086/311075}

\bibitem{Stella2001}
L.~{Stella}, X-RAY ASTRONOMY: Stellar Endpoints,AGN, and the Diffuse X-ray
  Background. \textbf{599}, 365 (2001).
\newblock \doi{10.1063/1.1434649}

\bibitem{Pappas2012b}
G.~{Pappas}, Mon. Not. Roy. Astron. Soc. \textbf{422}, 2581 (2012).
\newblock \doi{10.1111/j.1365-2966.2012.20817.x}

\bibitem{Maselli2015}
A.~{Maselli}, L.~{Gualtieri}, P.~{Pani}, L.~{Stella}, V.~{Ferrari}, ApJ
  \textbf{801}, 115 (2015).
\newblock \doi{10.1088/0004-637X/801/2/115}

\bibitem{Pappas2015b}
G.~Pappas,   (2015)

\bibitem{Klis2006}
M.~van~der Klis~in, \emph{Compact Stellar X-ray Sources (Cambridge
  Astrophysics)} (Cambridge University Press, 2006)

\bibitem{Rezzolla2003}
L.~{Rezzolla}, S.~{Yoshida}, T.J. {Maccarone}, O.~{Zanotti}, Mon. Not. Roy.
  Astron. Soc. \textbf{344}, L37 (2003).
\newblock \doi{10.1046/j.1365-8711.2003.07018.x}

\bibitem{Rezzolla2003a}
L.~{Rezzolla}, S.~{Yoshida}, O.~{Zanotti}, Mon. Not. Roy. Astron. Soc.
  \textbf{344}, 978 (2003).
\newblock \doi{10.1046/j.1365-8711.2003.07023.x}

\bibitem{Montero2004}
P.J. {Montero}, L.~{Rezzolla}, S.~{Yoshida}, Mon. Not. Roy. Astron. Soc.
  \textbf{354}, 1040 (2004).
\newblock \doi{10.1111/j.1365-2966.2004.08265.x}

\bibitem{Gondek-Rosinska2014a}
D.~{Gondek-Rosi{\'n}ska}, W.~{Klu{\'z}niak}, N.~{Stergioulas},
  M.~{Wi{\'s}niewicz}, Phys. Rev. D \textbf{89}(10), 104001 (2014).
\newblock \doi{10.1103/PhysRevD.89.104001}

\bibitem{DeDeo2004}
S.~{DeDeo}, D.~{Psaltis}, ArXiv Astrophysics e-prints  (2004)

\bibitem{Vincent2014}
F.H. {Vincent}, Classical and Quantum Gravity \textbf{31}(2), 025010 (2014).
\newblock \doi{10.1088/0264-9381/31/2/025010}

\bibitem{Doneva2014b}
D.D. Doneva, S.S. Yazadjiev, N.~Stergioulas, K.D. Kokkotas, T.M. Athanasiadis,
  Phys. Rev. D \textbf{90,}, 044004 (2014)

\bibitem{Ryan1995}
F.D. {Ryan}, Phys. Rev. D \textbf{52}, 5707 (1995).
\newblock \doi{10.1103/PhysRevD.52.5707}

\bibitem{Shibata1998}
M.~{Shibata}, M.~{Sasaki}, Phys. Rev. D \textbf{58}(10), 104011 (1998).
\newblock \doi{10.1103/PhysRevD.58.104011}

\bibitem{Pappas2012}
G.~{Pappas}, T.A. {Apostolatos}, Physical Review Letters \textbf{108}, 231104
  (2012).
\newblock \doi{10.1103/PhysRevLett.108.231104}

\bibitem{Yazadjiev2014}
S.S. Yazadjiev, D.D. Doneva, K.D. Kokkotas, K.V. Staykov, JCAP \textbf{1406},
  003 (2014).
\newblock \doi{10.1088/1475-7516/2014/06/003}

\bibitem{Staykov2014}
K.V. {Staykov}, D.D. {Doneva}, S.S. {Yazadjiev}, K.D. {Kokkotas}, JCAP
  \textbf{10}, 006 (2014).
\newblock \doi{10.1088/1475-7516/2014/10/006}

\bibitem{Yazadjiev2015}
S.S. {Yazadjiev}, D.D. {Doneva}, K.D. {Kokkotas}, Phys. Rev. D \textbf{91}(8),
  084018 (2015).
\newblock \doi{10.1103/PhysRevD.91.084018}

\bibitem{Read2009}
J.S. {Read}, B.D. {Lackey}, B.J. {Owen}, J.L. {Friedman}, Phys. Rev. D
  \textbf{79}(12), 124032 (2009).
\newblock \doi{10.1103/PhysRevD.79.124032}

\bibitem{Gondek-Rosinska2008}
D.~{Gondek-Rosinska}, F.~{Limousin}, ArXiv e-prints  (2008)

\bibitem{Naef2010}
J.~{N{\"a}f}, P.~{Jetzer}, Phys. Rev. D \textbf{81}(10), 104003 (2010).
\newblock \doi{10.1103/PhysRevD.81.104003}

\end{thebibliography}

\end{document}